# Generation of Efficient Codes for Realizing Boolean Functions in Nanotechnologies

**Ashish K. Singh, Adnan Aziz, Michael Orshansky, Sriram Vishwanath**

**Department of Electrical and Computer Engineering, University of Texas at Austin**

## 1. Abstract

We address the challenge of implementing reliable computation of Boolean functions in future nanocircuit fabrics. Such fabrics are projected to have very high defect rates. We overcome this limitation by using a combination of cheap but unreliable nanodevices and reliable but expensive CMOS devices. In our approach, defect tolerance is achieved through a novel coding of Boolean functions; specifically, we exploit the don't cares of Boolean functions encountered in multi-level Boolean logic networks for constructing better codes. We show that compared to direct application of existing coding techniques, the coding overhead in terms of extra bits can be reduced, on average by 23%, and savings can go up to 34%. We demonstrate that by incorporating efficient coding techniques more than a 40% average yield improvement is possible in case of 1% and 0.1% defect rates. With 0.1% defect density, the savings can be up to 90%.

## 2. Introduction

The end of CMOS scaling will require finding alternatives to CMOS electronics. The currently explored technologies include carbon nanotube (CNT), molecular, and quantum devices [1]. It is premature to conclude which particular device technology will ultimately prove to be most promising, although a technology based on FETs with carbon nanotubes as channels appears to be the most likely candidate for near-term acceptance [2][3].

The extreme *defect densities* and high *parametric variability* in emerging device technologies pose a fundamental new challenge which requires researching new cost-effective strategies for ensuring reliable computation [1]. Reliable computation in integrated digital circuits has been a subject of extensive research. A common paradigm for fault tolerant system design is the use of *redundancy*, such as triple modular redundancy (TMR), which uses three copies of a computational unit and arbiters employing majority voting to produce the correct value, or reconfiguration [4]. Redundancy combats both permanent and transient faults, but reconfiguration helps deal only with permanent faults.

Most of the existing work is not directly relevant to finding a solution for the challenges of the future device technologies**:** (1) With the low defect densities, it was reasonable to assume during the redundancy insertion that the probability of the arbiter failing was negligible. This assumption becomes unreasonable in the case of nanotechnology, endangering the entire traditional fault tolerance edifice. (2) The traditional reconfiguration approach also fails because, with device defect density being of the order of 1-5%, the probability that a module which contains more than a few gates will operate correctly becomes so low that orders of magnitude more spares than actual working components are needed. This is in contrast to the traditional silicon technologies, in which low defect density requires relatively few spares and allows performing reconfiguration at the module level [4].

At higher defect densities, defect tolerance must be addressed at a lower level of design, specifically, at the level of logic gates. The need to make individual gates reliable changes the traditional approach to fault tolerance in a fundamental manner. The reason is that now the complexity of the test and reconfiguration circuitry, as well as the arbitration circuitry becomes comparable to that of logic itself. Consequently, we believe that the right design style is to make individual modules more reliable by using redundancy internal to each module, and to choose an optimal combination of inter-module reconfigurability and intra-module redundancy.

Just as communication in the presence of noise can be made reliable by adding redundant code bits, logic circuits can be made reliable out of noisy gates by the use of redundant gates. The possibility of building reliable circuits from noisy gates has been studied theoretically, most famously, by von Neumann [6]. He introduced a massively redundant strategy, now known as von Neumann or NAND multiplexing. He demonstrated that, at the cost of tremendous overhead, for any Boolean function a reliable circuit can be synthesized even if individual devices fail with a fairly high probability (his bound was 11.87%). His construction inserted an exponential amount of redundant logic for all functions. Later, Pippenger refined von Neumann's results, showing that for a functional selected at random, the coding overhead was a constant factor, with high probability [7]; however, Pippenger's construction is exponential for the functions encountered in practice. In summary, these works have shown that it is possible to construct reliable circuits from individually unreliable gates, but the cost is prohibitively high, resulting in the area overhead of $10^3$-$10^4$ of the original circuit [8]. Because of that, gate-level redundancy insertion for constructing reliable logic circuits has not been previously exploited in VLSI design and remained a theoretical possibility.

In this paper we propose a design methodology for building reliable computational elements using devices manufactured in nano-technologies. The fundamental strategy of fault-tolerance is not reconfiguration, but low-level protection of individual Boolean functions through the use of efficient coding. The Boolean functions are represented in terms of ROMs or truth tables, and a novel version of the Hamming code is used to protect the functions. The proposed coding strategy exploits for the first time the structure of Boolean logic networks to produce better codes

The proposed computational architecture is based on a heterogeneous CMOS - carbon nanotube fabric. It seems certain that the early practical uses of CNT-based electronics will be built on top of the heterogeneous CMOS-CNT processes [9]. The enormous economic investments into the CMOS technology and design infrastructure provide a strong impetus to leverage the existing flows as much as possible. Interfacing CMOS and CNT technology is technically feasible; the economic viability of integration is seen by the recent commercialization of non-volatile memory circuits based on CNT integrated with CMOS. Our approach is predicated on optimally combining reliable, albeit low performance, CMOS components with high performance, albeit unreliable, CNT devices.

The remainder of the paper is organized as follows. In Section 3, we describe the architecture of the heterogeneous circuit fabric in which the proposed coded Boolean functions can be decoded for achieving error correction. In Section 4 we describe the traditional Hamming code being applied for error correction of Boolean functions. Section 5 describes the use of don't cares for reduction of redundancy required for coded Boolean functions. Section 6 gives a formal description of code construction and minimization of overhead which is solved by converting the problem to the Boolean satisfiability problem. Experiments for redundancy reduction and yield enhancement due to error correction are presented in Section 7, and in section 8 we summarize the key contributions.

# 3. Heterogeneous Fabric and Architecture

In building a reliable circuit from unreliable gates, we face a fundamental limitation that the overall reliability is limited by that of the gate driving the output (typically an arbiter). Enhancing the reliability of arbiters would greatly reduce the complexity of building reliable circuits. The heterogeneous fabric resolves this: CNT devices can be used as the base "noisy" logic which naturally exploits its advantages (low-cost high-density, low power); and CMOS gates can be used for recovery of correct response, i.e., for decoding, by utilizing a much higher reliability of silicon devices. One simple way of mixing CMOS and nanodevices is to build *c* independent copies of circuits for each block, and use a CMOS majority voting element. This would require at least triplication of the nano-based computing blocks. We propose here a different approach that fundamentally relies on sophisticated coding.

The optimal size of nano-based ROMs and CMOS-based decoding logic is dependent on a number of technology and circuit characteristics of both fabrics. Despite the lack of clarity about CNT-based electronics, some first order estimates based on experimental results can already be made. Here we compare the key CNT properties to that of a CMOS technology at the end of the roadmap, 22nm node. The projected CMOS transistor density is from $1.2 \cdot 10^9$/cm2 for logic to $24 \cdot 10^9$/cm2 for DRAM. This is compared to substantially higher density of $10^{12}$/cm2 that has been reported for a CNT-based memory array [1]. The switching time of a NMOS device in a 22nm CMOS is projected to be 0.15ps, which, surprisingly, appears to be comparable to the CNT switching time of 0.5-1ps. A significant difference can be seen in the switching energy of an NMOS device (0.004 fJ/device) and the estimated CNT energy per switching ranging from 0.0005fJ/device to 0.02fJ/device. From the above analysis, it appears that the major advantage of CNT-based circuitry over the CMOS-based circuitry is in the transistor density and switching energy, with a difference of at least a factor of 40 for density and up to 10x for switching energy. Thus, the optimal granularity of inserting CMOS-based arbiters will be limited not by the timing overhead but by the density and power constraints.

We propose to partition the logic into blocks, which individually can be implemented by a lookup-table (LUT), e.g., a ROM. The basic logic blocks will be "protected" using a coding scheme described in the next sections. By using coding at a very low level of granularity, our technique enables a much higher probability of instantiation of a block in a defective fabric. We specifically address the challenge of protecting the content of memory bits, which are taken throughout the paper to be the basic units of data storage in a ROM, such as in a NAND or NOR ROM architectures. The row address decoders can be protected through the use of a simple coding technique [11]. The decoder is implemented in CMOS and thus imposes area overhead compared to dense nano-LUTs. The overhead can be reduced by sharing the decoders using the time multiplexing strategy. We envision that the combinational circuit is evaluated in a sequence of microcycles, e.g., Chapter 6 of [18]. This is illustrated in Figure 1. The starting combinational circuit has four multi-input multi-ouput logic blocks. We apply our coding to each block, to produce the functions CF0-CF3. The output function is now computed by first decoding the output of CF0 and storing the result in a register. We then compute the output of CF1, and store it in a register. The inputs to CF2 are now decoded, so we take its output and decode it, storing the result in the third register. Now the inputs to CF3 are decoded, and the decoder output is the final output. There are register re-use and scheduler cost issues which will have to be resolved, and in this paper, these challenges are not addressed in any comprehensive manner.

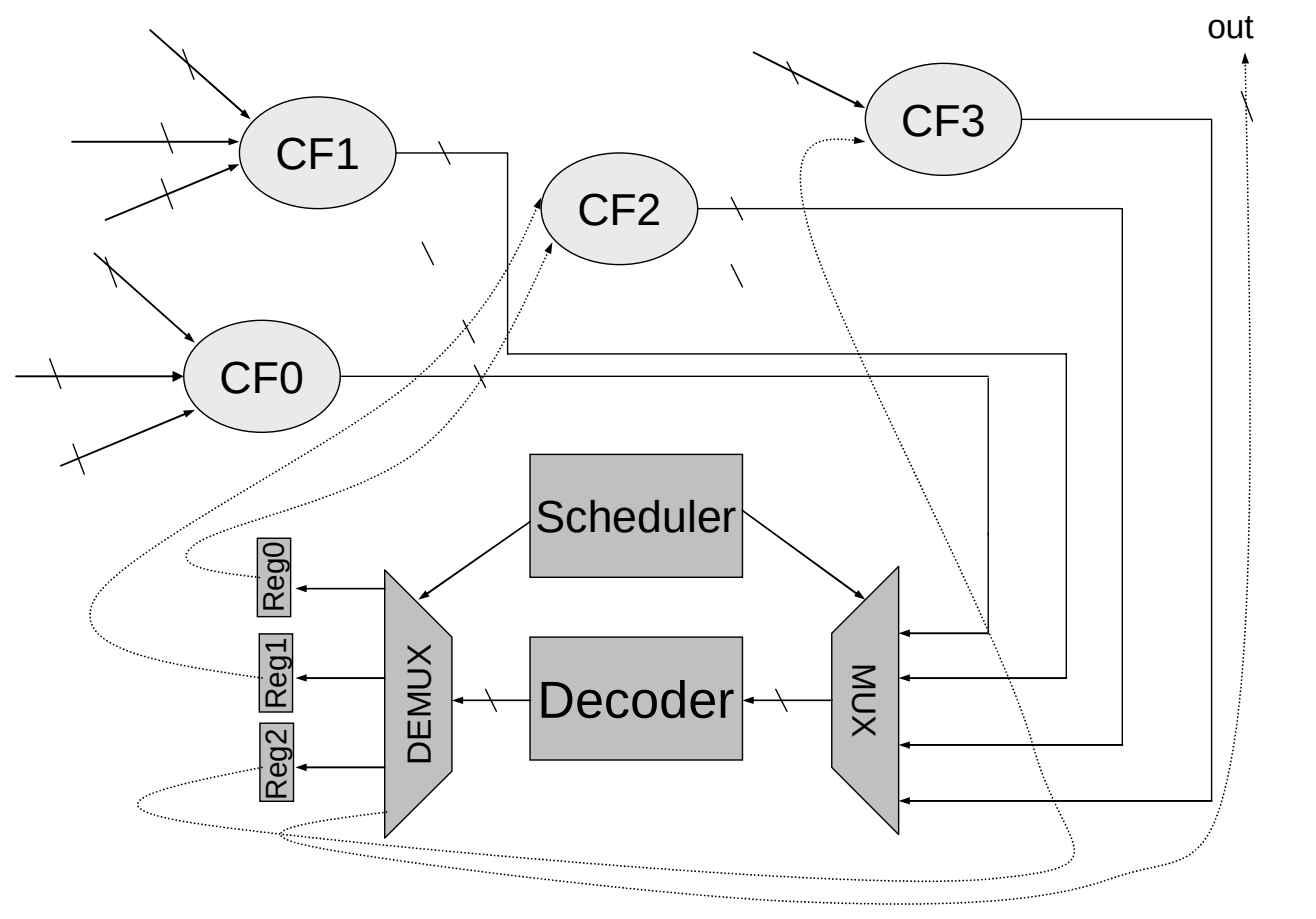

**Figure 1. Microsequenced logic.**

# 4. Coding of Logic Functions

Coding theory has been originally developed for communication. Codes are used to prevent corruption of signals during transmission over a channel. A key distinction between a channel and a logic circuit is that the former simply transmits bits, whereas the latter transforms their values. The transformative nature of logic operations makes the extension of coding for circuits very difficult. It has been proven that for 2-input Boolean functions other than XOR/XNOR, the best codes are repetition codes; however, this is not generally the case for more complex networks [14]. Repetition codes require duplication for error-detection, and triplication for error correction. From the view point that is more familiar to circuit designers, repetition coding is equivalent to triple modular redundancy (TMR).

In this paper, we advocate a novel approach to coding for logic. It is based on representing Boolean computation in terms of ROM-based logic. ROMs are universal computing devices that can implement any logical function. The motivation for adopting this representation is the existence of codes for spatial channels, i.e., memories, which we can exploit. Coding for memories has received significant attention in the information and coding theory communities. The study of the fundamental limits on the number of bits that can be stored in a memory with defects was pioneered by [12], with the capacity of classes of these systems determined in [13]. However, much of the prior work in this field has usually been more analytic in nature without concrete designs of codes and decoding algorithms to go with this analysis.

The design of the decoder depends on the code selected. In general, the decoder proceeds by iteratively making small changes to the word being decoded, stopping when it is a legal codeword [10]. Because the cost of decoding is important for making the proposed flow practical, we will be constraining ourselves to linear codes for which decoders are easier to build. Linear error correction code design is the art of determining appropriate subspaces of n-dimensional vector spaces with special properties. The Hamming code and its shortened versions have many special properties making them particularly suitable for nano-circuits. First, they are *perfect codes* [10] in that they can always correct exactly one error in the input bit-words. They also possess a highly intuitive structure and an elegant and simple decoding algorithm.

We propose a coding scheme for general Boolean functions, i.e., functions with n-inputs and m-outputs. Multi-output functions can be identified in multi-level logic networks [19]. We propose a coding scheme for general Boolean functions, i.e., functions with n-inputs and m-outputs. This scheme is based on the Hamming code – a minimum number of redundant outputs are added as parity bits for coding. As is the case in coding for communication, the crucial consideration is the overhead of coding – the number of bits that need to be added. The Hamming code is optimal; a typical Hamming code is $(2^m-1, 2^m-m-1)$, in other words, for $2^m-m-1$ data bits, $m$ parity bits need to be added for full protection. A single decoder will produce a correct output for any input bitword under one bit error scenario.

Consider a Boolean function with 2 inputs $(x_0, x_1)$ and 3 outputs $(y_0,\ y_1,\ y_2)$. The outputs are defined as $y_0 = x_0 \vee x_1$, $y_1 = x_0 \vee x_1$ and $y_2 = x_0 + \bar{x}_1$. The truth table is shown below

| $x_0\ x_1$ | $y_0\ y_1\ y_2$ |
|---|---|
| 00 | 001 |
| 01 | 110 |
| 10 | 111 |
| 11 | 111 |

The standard Hamming decoder can be built for the above data by adding three redundant columns as shown below:

$$\underbrace{\begin{bmatrix} 0 & 0 & 1 & 0 & 1 & 1 \\ 1 & 1 & 0 & 0 & 1 & 1 \\ 1 & 1 & 1 & 0 & 0 & 0 \\ 1 & 1 & 1 & 0 & 0 & 0 \end{bmatrix}}_{\substack{\text{Original LUT rows +} \\ \text{3 redundant columns}}} \times \underbrace{\begin{bmatrix} 0 & 1 & 1 \\ 1 & 0 & 1 \\ 1 & 1 & 0 \\ 0 & 0 & 1 \\ 0 & 1 & 0 \\ 1 & 0 & 0 \end{bmatrix}}_{\text{Decoder_Matrix}^{\text{T}}} = [0]_{4\times 7}$$

The last three columns essentially represent the parity bits of $\{y_0, y_1\}$, $\{y_0, y_2\}$ and $\{y_1, y_2\}$ output bits respectively. The absence of an error can be detected by multiplying the output row with the transpose of decoder matrix and testing if the product is a 0-row vector. Otherwise, in a case of single bit error, one can compute the product of erroneous row with the transpose of the decoder matrix to get a row vector known as *syndrome*. The Hamming codes are special precisely because of the existence of simple syndromes: if there is an error in bit location $i$, then the syndrome will be identical to the *i*-th row. The cost of coding in this case is 3 extra columns added to an LUT. It is important to reduce this cost.

An important contribution that this work makes is the extension of earlier Hamming code constructions. In the next section, we exploit the fact that Boolean functions, and thus LUTs, may contain don't cares, to achieve reduction in the number of redundant columns required.

# 5. Using Don't Cares for Compact Coding

All earlier applications of coding in fault-tolerant memory design have treated the pattern to be protected as simply a set of 1s and 0s. Here we show how the structure of Boolean functions can be used to reduce the coding overhead.

A Boolean function is defined by its ON-set, OFF-set, and its DC-set [16]. DC-set (don't-care set) is a set of inputs on which the output can be *either* 0 or 1. In control logic, many practically used functions are defined with a fairly compact ON-set, giving rise to a large DC-set. There are multiple ways to represent don't-care sets. In developing a technique for coding logic, we will rely on the LUT-based representation of Boolean functions. LUTs are essentially truth tables, and the DC-sets are captured here explicitly. For multi-level circuits, local don't care sets can be computed using standard existing techniques [16].

*The key observation that we make is that it is possible to assign values to the members of the don't care set, so as to encode the logic function more compactly.* Suppose we have some don't cares in the function as shown below.

| $x_0\ x_1$ | $y_0\ y_1\ y_2$ |
|---|---|
| 00 | 00X |
| 01 | 11X |
| 10 | 111 |
| 11 | XX1 |

Our task is to find optimal assignments to the don't cares so as to minimize the number of redundant columns. In this case there are 4 don't cares remaining, which may be assigned so that we need 0 redundant columns. If we assign the first don't care to 0 and the last three to 1, then we can construct a decoding matrix without adding any redundant columns as shown below:

$$\underbrace{\begin{bmatrix} 0 & 0 & 0 \\ 1 & 1 & 1 \\ 1 & 1 & 1 \\ 1 & 1 & 1 \end{bmatrix}}_{\text{Original LUT rows}} \times \underbrace{\begin{bmatrix} 0 & 1 \\ 1 & 0 \\ 1 & 1 \end{bmatrix}}_{\text{Decoder_Matrix}^{\text{T}}} = [0]_{4\times 2}$$

For this example, we can correct single bit errors without any additional outputs because the only two legal outputs are 000 and 111, which have a Hamming distance of 3. The ability to reduce the cost of coding in this way is the basic intuition behind our work.

In order to exploit the existence of don't care for more compact coding, an efficient algorithm for constructing such codes needs to be developed. It follows from a reduction from graph coloring that the optimization problem for forming such codes is NP-complete.

Below we describe an algorithm that casts the problem as CNF-SAT, and solves it efficiently using a SAT solver, e.g., MiniSat [17].

# 6. Efficient Coding using Don't Care Conditions

We use a single error correcting Hamming code on each row of the LUT. Experimental results support the sufficiency of single bit error correction for realistic defect densities. A traditional Hamming code for correcting $n$-bit original data requires $s$ extra *parity bits*, where $s$ the least integer is such that $2^s \geq n+s$. The default number of extra bits required for Hamming code construction is denoted as $\triangle(n)$ .One contribution of the work is *reduction* of the number of default redundant columns that one need in the traditional Hamming code by exploiting the actual data present in LUT and presence of don't cares.

Let $D_{l\times p}$ be a known Boolean matrix consisting of entries in certain LUT with $l$ rows and $p$ outputs. The Hamming code is described by a *decoder matrix* $H_{\lceil \log_2 p\rceil \times p}$. The decoding procedure is simple and based on the given decoder matrix. Suppose the output data which correspond to the *i*-th row of $D_{l\times p}$ has a single bit error. Let this row vector be denotes as $D_{(i)}$ and location of error bit be *j*-th starting from the left. Let the *j*-th column of $H$ be denoted as $H_{[j]}$ then the following identity can be established which computes a row vector known as *syndrome*

(1) $$D_{(i)}.H^T = H_{[j]}{}^T$$

If the syndrome vector contains 0s in all the entries, there is no error. If there is a single bit error in location $j$ of the data then, the syndrome will be equal to the $j_{th}$ column of $H$ which has a one to one mapping to the bit at position $j$. This allows us to uniquely identify the error and then fix it by inverting erroneous bit.

The code construction is essentially the construction of the decoder matrix *H.* In general, the code construction requires adding additional parity bits to data. When $D$ contains exhaustively all possible Boolean vector of size $p$ as its rows, we always need to add $\triangle(p)$ number of parity bits. However, for some $D_{l\times p}$ , such as in the example previously studied, it is possible that a decoding matrix exists even without adding extra bits. We want to exploit this for constructing compact codes. Whether or not $D_{l\times p}$ requires extra parity columns can be determined by checking the following condition.

Define $t$ to be the smallest number such that $2^t > p$. There exists a matrix $H_{t\times p}$ such that no two columns are identical and no single column of $H$ be identically zero. Furthermore, the following condition holds $DH^T = [0]_{l\times t}$. If these conditions hold, we can design a decoder, without adding any redundant columns, which will correct a single bit error in each row of the data.

If the above condition is not met, some minimum extra columns need to be added to $D$ , until the condition $DH^T = [0]_{l\times t}$ is satisfied for some $H$ . Since we want to add the smallest number of extra bits, we can test the condition after adding each new column. The existence of don't cares allows greater flexibility in finding a solution to the above existence check. By proper assignment of don't cares, we increase the likelihood that fewer redundant column will be needed.

The algorithm for identification of the decoding matrix is mathematically formulated as a *Boolean satisfiability problem* which is efficiently solved using a SAT solver [15]. This formulation also permits seamless handling of don't cares in the matrix

We convert $DH^T = [0]_{l\times t}$ to the following equivalent form

(2) $$\sum_{j=1}^{p} d_{ij}.h_{kj} \equiv 0 \qquad \forall i \in [1,l] \text{ and } k \in [1,t]$$

This problem can be easily converted to a Boolean problem as follows: ( $\otimes$, $\wedge$ denotes XOR and AND of two Boolean variables)

(3) $$\mathop{\otimes}_{j=1}^{p} (d_{ij} \wedge h_{kj}) \equiv 0$$

Now using Tstein-transformation [15] we can convert (3) into a CNF form efficiently by introducing a small overhead of extra variables. The equivalent CNF can be checked for satisfiability using a standard SAT solver [15]. The further constrains on the matrix $H$ are that no column should be identically 0 and no two columns should be identical can be easily turned into Boolean constraints as follows

(4) $$\mathop{\vee}_{i=1}^{t} h_{ij} \equiv 1 \quad \forall j \in [1,p]$$
$$\mathop{\vee}_{i=1}^{t} (h_{ij_1} \otimes h_{ij_2}) \equiv 1 \quad \forall j_1, j_2 \in [1,p], j_1 \neq j_2$$

These constraints can again be converted into CNF format, using the Tstein-transformation and given to the SAT solver along with (3). If the SAT solver detects that the set of Boolean conditions are satisfiable for some assignment to the entries of decoder matrix $H$ , don't cares and redundant column entries of $D$ then a feasible decoding matrix exists and we can find the entries of $H$ from SAT solution.

The number of variables and clauses required for SAT problem is dependent on the dimension of the data matrix $D$ , and it can be summarized as: variables including the extra variable due to Tstein transformation $(4p-2)lt + p + {}^pC_2(4t+1)$ clauses required: $tp + (p-1)lt + {}^pC_2 t$ . Thus both the number of clauses and the number of rows are linear in $l$ (the number of rows in $D$ ). Since $t = O(\log_2 p)$ and $p$ is small there is not a significant impact on the resulting CNF problem due to $p$ and $t$ .

If the SAT problem is unsatisfiable, we need to add a certain number of redundant numbers of columns in matrix $D$ so as to satisfy the condition in (2) and conditions on the decoder matrix (4). Suppose we wanted to test if by adding $k$ extra columns to $D$ we can get a feasible decoder matrix. This will only cause the dimensions of $D$ and $H$ to change and the Boolean conditions presented in (3), (4) remains valid and can be applied to get the new decoder matrix $H$ as well as the redundant column entries.

If in addition $D$ contains some don't cares, this will increase the likelihood of redundant column reduction, by proper assignment of don't cares. The assignment problem for don't cares has been incorporated in the framework of the formulation presented before. We can always get a feasible decoder matrix by adding $\triangle(p)$ redundant columns using traditional hamming code method according to the lemma stated below:

**Lemma**: There exists a matrix $X_{l\times\triangle(p)}$ such that $[D\ X]_{l\times(p+\triangle(p))}$ admits a feasible decode matrix $H_{\triangle(p)\times(p+\triangle(p))}$. The matrix $H$ is constructed by keeping in its columns $\triangle(p)$ bit long Boolean representation of the numbers starting from 1 to $(p+\triangle(p))$ with the power of 2 coming in end in increasing order and the rest of the numbers coming before also in increasing order.

Thus we start in an incremental fashion by adding 1 extra redundant column and checking the feasibility of decode matrix by solving (2). The procedure is guaranteed to terminate by Lemma stated before. Since $\triangle(p) \sim O(\ln p)$, the run-time overhead is small for this iterative approach.

# 7. Experiments and Probabilistic Analysis for Yield Improvement

We first studied the extent to which it is possible to reduce the addition of redundant columns in code construction. For this setup, we considered LUTs of sizes $2^3$x3 (3 inputs, 3 outputs) and $2^4$x4 (4 inputs, 4 outputs). The experiments were conducted using 5000 randomly generated 0-1 matrices. We considered two scenarios: "symmetric" - in which the probability of 0 and 1 were equal, and "skewed" - in which the probability of 1 was 0.2. The experiments were repeated after introducing don't-care conditions into LUTs, with the number of don't cares set to 50% of all entries, which seems to be a reasonable number.

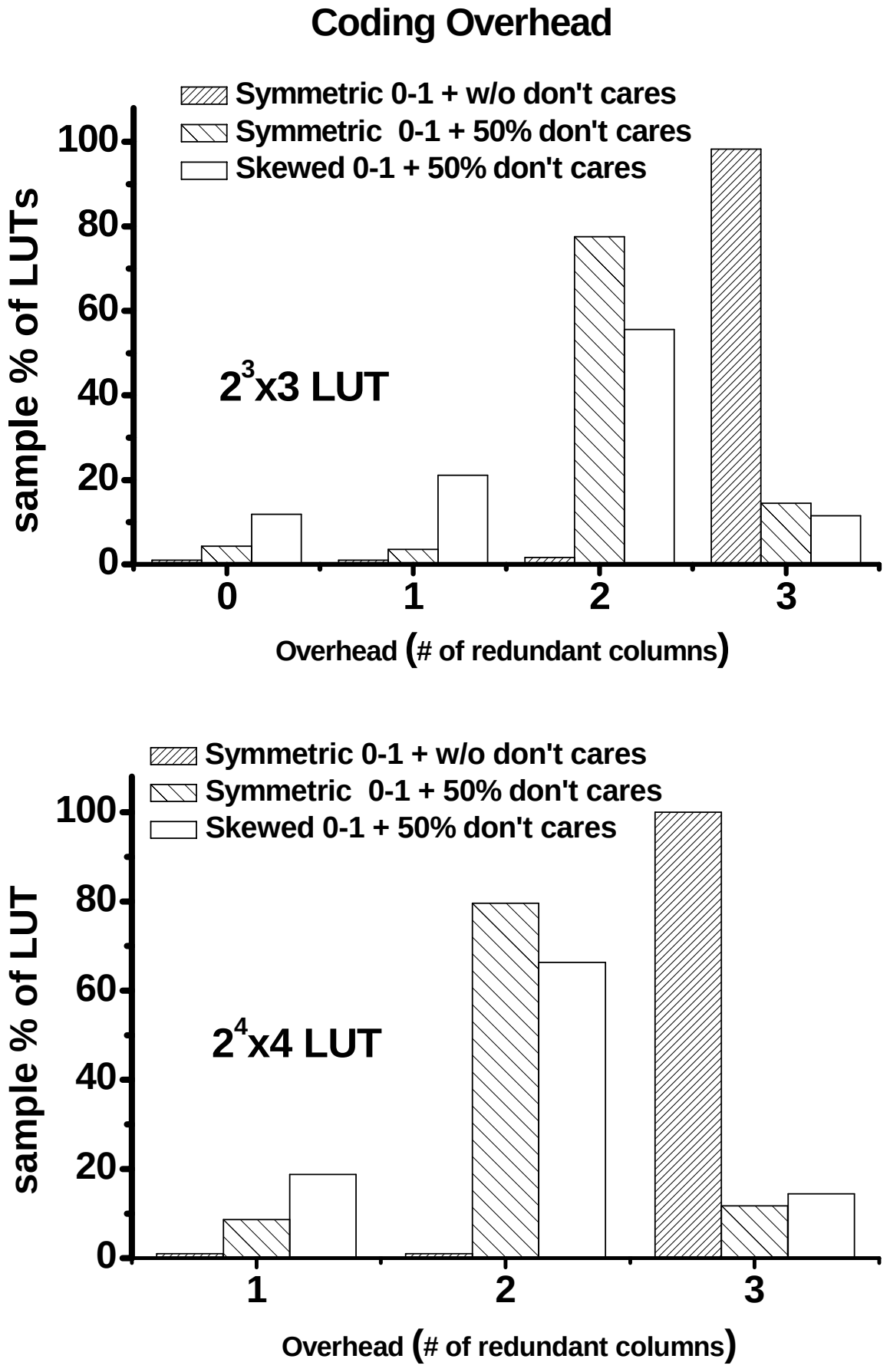

**Figure 2. Coding overhead in terms of extra columns. Don't cares significantly increase fraction of LUTs requiring smaller overhead.**

The default number of extra bits required by the Hamming code is 3 for both matrices. Figure 2 shows that in the case of $16\times4$ LUT, almost all LUTs required the maximal number of redundant columns 3 in the absence of don't cares. Notably, introducing the don't cares allowed 80% of all LUTs to require only 2 redundant columns. Experiments also indicate that symmetric matrices are more difficult to encode - skewed LUTs permitted a higher reduction in the number of redundant columns. We estimated the reduction of area compared to naïve coding by defining the area of an LUT to be the number of bits in it that includes the redundant bits. The use of don't cares allows us to reduce the area by 23%, on average, and savings range from 16% to 34%.

We also studied improvements enabled by the single bit error correction. Consider an LUT with $m$ rows and $n$ columns. Let $p$ be the *defect probability* of a single bit. Let the defect probability of individual bits be independent. A single row consisting of $n$ bits is error free with probability of $(1-p)^n$. Now if there are $f(n)$ rows of size $n$ in a circuit, then the probability of the error free circuit is $\psi(n)=(1-p)^{nf(n)}$. Suppose we include single bit error correction and add $s(n)$ redundant bits. In the worst case, $s(n)$ is always bounded by $\triangle(n)$. For the case of single bit error correction, the probability of an error free single row of n-bits becomes

$$(5)\qquad \chi(n)=(1-p)^{n+s(n)}+(n+s(n)-1)(1-p)^{n+s(n)-1}p$$

For the entire circuit the error free probability is $\chi(n)^{f(n)}$. In general, we may have LUTs with different numbers of outputs; thus the overall yield probability will be the product of $\psi(n)$ or $\chi(n)$ over all feasible $n$ in the case of no error correction, and single error correction, respectively.

Assuming that we always use the default number of redundant columns, we estimate the yield for the LUT with outputs of size 3 and 4. We assume that there are $2^{16}$ blocks where each block is a single LUT with 16 rows. We varied the bit failure probability from 1e-5 to 16e-5. The yield with no error correction remained smaller that 1e-9 in the all the cases. With our single-bit error correction, yield remained above 70% for most of the cases, Figure 3. Without error correction, reasonable yield can be expected for defect densities that would have to be 1000X smaller. In another experiment we varied single bit failure probability from 1e-8 to 25e-8. In this range of failure probabilities the yield with a single bit error correction remains always above 99%, Figure 4.

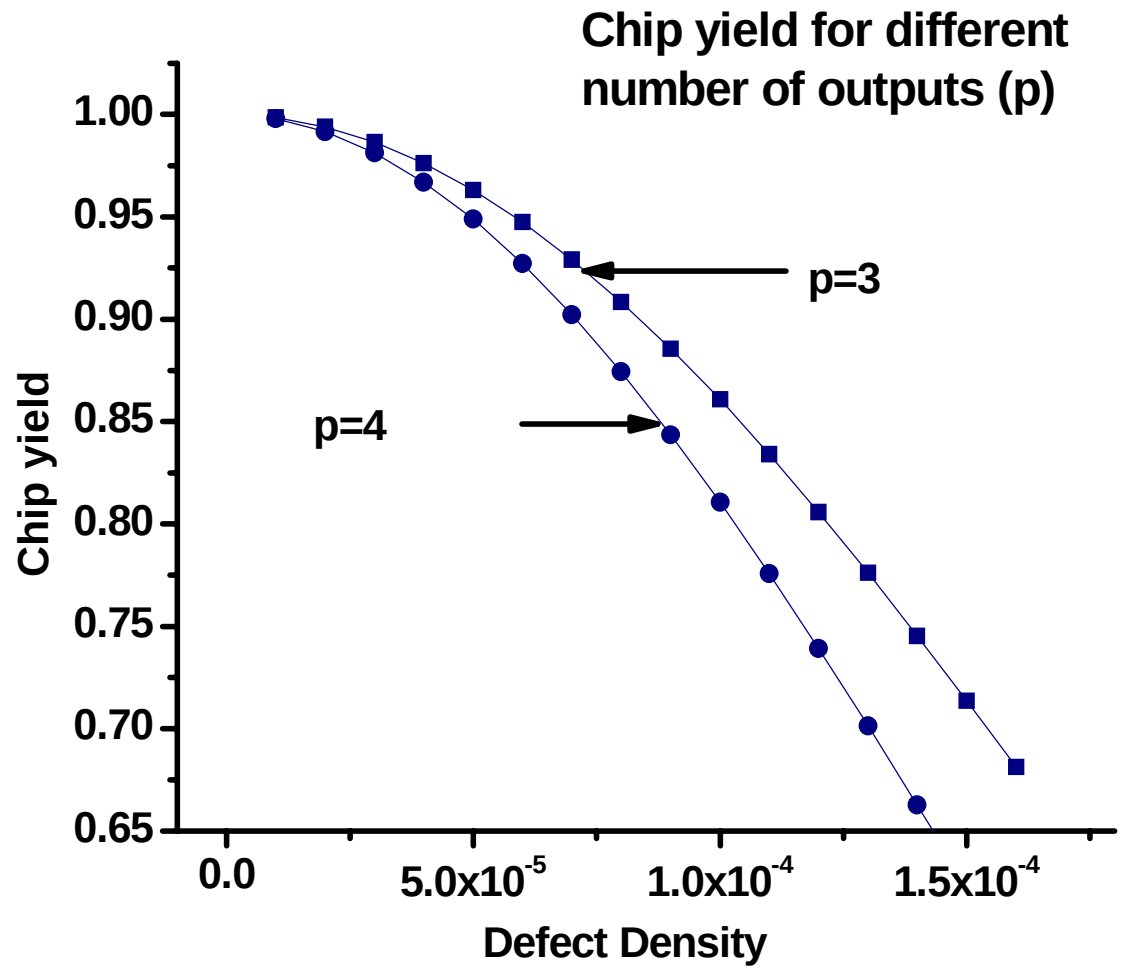

**Figure 3. Our coding strategy allows achieving good chip yields. Yield without error correction is effectively 0. The number of LUTs is $2^{16}$.**

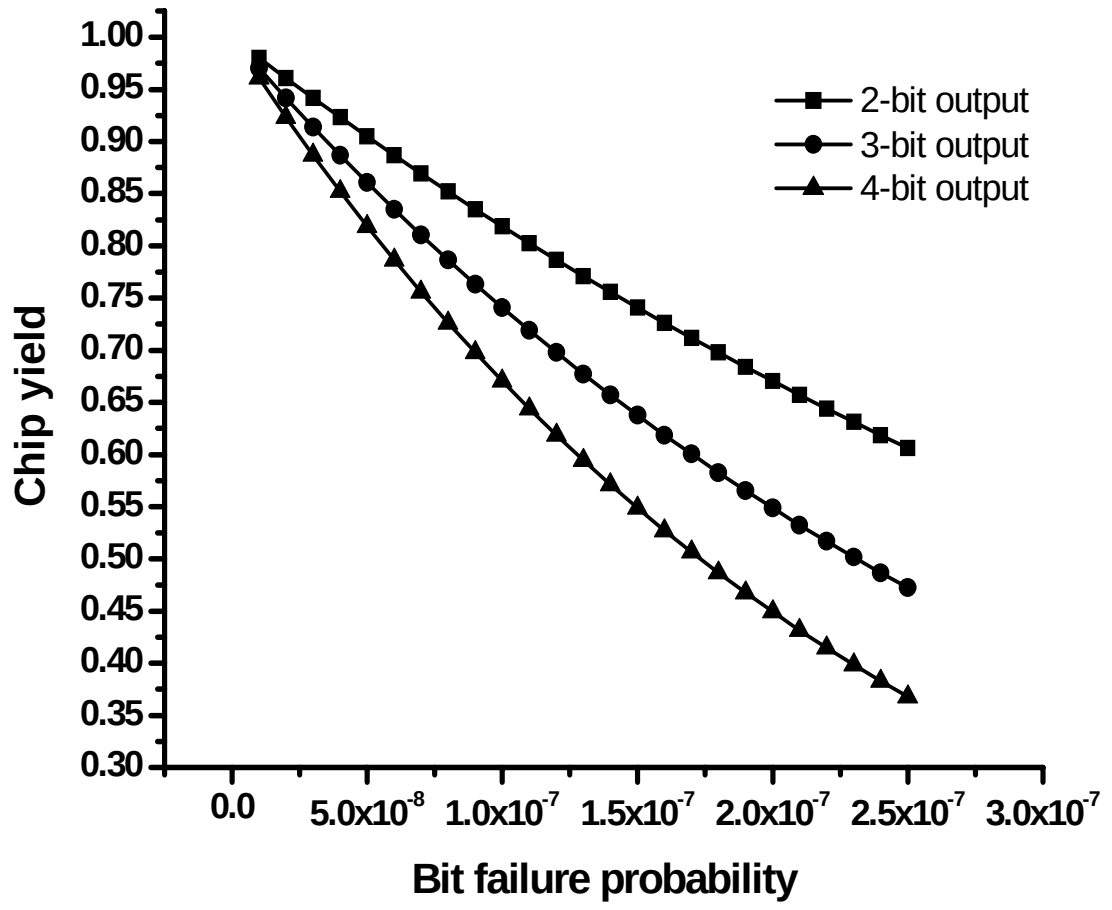

**Figure 4. Yield without any error correction. For the range of failure probability shown, yield with a single bit error correction remained above 99%.**

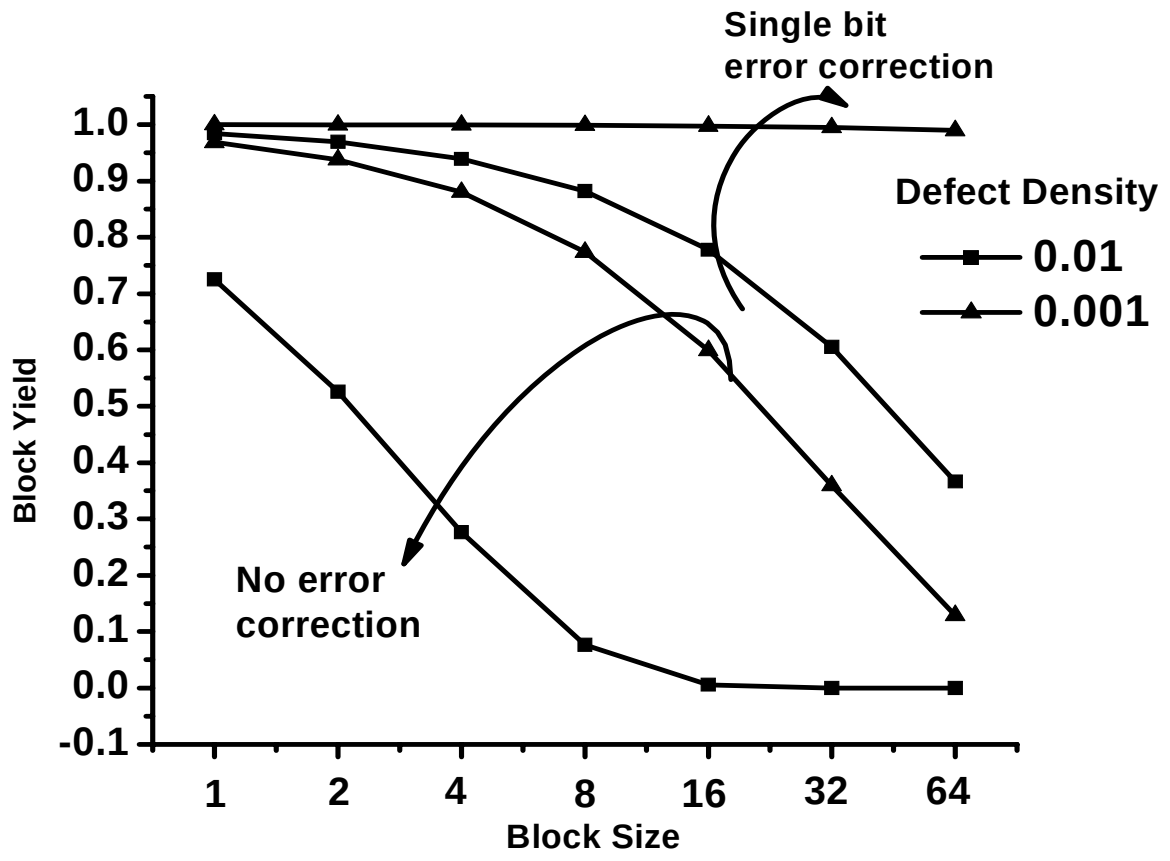

**Figure 5. Our error correction scheme raises significantly the block yield, i.e. the probability of instantiation even for very high defect density (.01).**

Finally, we studied the amount of yield improvement permitted by our error correction for different block sizes and defect densities. We considered fairly high defect densities of 0.01 and 0.001. Such defect densities (probabilities) may be expected in nano-technologies. The results are for LUTs with 2 outputs, Figure 5. We can observe that for defect density of 0.01, the yield in the case of single-bit error correction dropped from 99% to 40% with block size increase. If we do not apply error correction, yield quickly drops to nearly 0% yield when the number of blocks increases to 8. The yield improvement for block of sizes of 16 and more is higher than 30%. For the defect density of 0.001, the yield with a single bit error correction remained more than 99% across all block sizes while the yield without error correction drops to less than 10%.

# 8. Conclusion

We introduced a methodology for realizing coded Boolean functions implemented in nano-gates. We showed that this coded implementation along with the CMOS decoder allows us to increase the yield of nano-circuits significantly in presence of high defect density. We propose novel extensions to the Hamming code techniques to reduce the coding overhead using the special structure of Boolean functions and the presence of don’t cares. We also described a heterogeneous circuit fabric and architecture in which such a coded implementation of Boolean function can be effectively realized.